\documentclass[prl,nofootinbib,aps,twocolumn,preprintnumbers,amsmath,amssymb,floatfix,superscriptaddress]{revtex4-1}
\usepackage{graphicx}
\usepackage{xspace}
\usepackage{booktabs}
\usepackage{bbm}
\usepackage{color}
\usepackage[colorlinks=true,linkcolor=blue,urlcolor=magenta,citecolor=blue]{hyperref}

\newcommand{\eV}{\mathrm{eV}}

\newcommand{\be}{\begin{equation}}
\newcommand{\ee}{\end{equation}}
\newcommand{\benn}{\begin{equation*}}
\newcommand{\eenn}{\end{equation*}}
\newcommand{\bse}{\begin{subequations}}
\newcommand{\ese}{\end{subequations}}

\renewcommand\({\left(}
\renewcommand\){\right)}

\newcommand{\exclude}[1]{}

\begin{document}
%
%
\preprint{MPP-2016-141}
%
%
%
\title{Dielectric Haloscopes: A New Way to Detect Axion Dark Matter}

\author{Allen~Caldwell}
\affiliation{Max-Planck-Institut f\"ur Physik (Werner-Heisenberg-Institut),
  F\"ohringer Ring 6, 80805 M\"unchen, Germany}

\author{Gia~Dvali}
\affiliation{Max-Planck-Institut f\"ur Physik (Werner-Heisenberg-Institut),
F\"ohringer Ring 6, 80805 M\"unchen, Germany}
\affiliation{Ludwig-Maximilians-Universit\"at,
Theresienstra{\ss}e 37, 80333 M\"unchen, Germany}
\affiliation{CCPP, New York University, New York, NY 10003, USA}

\author{B\'ela~Majorovits}
\affiliation{Max-Planck-Institut f\"ur Physik (Werner-Heisenberg-Institut),
  F\"ohringer Ring 6, 80805 M\"unchen, Germany}

\author{Alexander~Millar}
\affiliation{Max-Planck-Institut f\"ur Physik (Werner-Heisenberg-Institut),
F\"ohringer Ring 6, 80805 M\"unchen, Germany}

\author{Georg~Raffelt}
\affiliation{Max-Planck-Institut f\"ur Physik (Werner-Heisenberg-Institut),
  F\"ohringer Ring 6, 80805 M\"unchen, Germany}

\author{Javier~Redondo}
\affiliation{Max-Planck-Institut f\"ur Physik (Werner-Heisenberg-Institut),
F\"ohringer Ring 6, 80805 M\"unchen, Germany}
\affiliation{University of Zaragoza,
P.\ Cerbuna 12, 50009 Zaragoza, Spain}

\author{Olaf~Reimann}
\affiliation{Max-Planck-Institut f\"ur Physik (Werner-Heisenberg-Institut),
  F\"ohringer Ring 6, 80805 M\"unchen, Germany}

\author{Frank~Simon}
\affiliation{Max-Planck-Institut f\"ur Physik (Werner-Heisenberg-Institut),
F\"ohringer Ring 6, 80805 M\"unchen, Germany}

\author{Frank~Steffen}
\affiliation{Max-Planck-Institut f\"ur Physik (Werner-Heisenberg-Institut),
  F\"ohringer Ring 6, 80805 M\"unchen, Germany}

\collaboration{The MADMAX Working Group}

\begin{abstract}
	We propose a new strategy to search for dark matter axions in the mass range of 40--400\,$\mu\eV$
	by introducing dielectric haloscopes, which consist of dielectric disks placed in a magnetic field.
	The changing dielectric media cause discontinuities in the axion-induced electric field,
	leading to the generation of propagating electromagnetic waves to satisfy the continuity requirements at the interfaces. Large-area disks with adjustable distances 
    boost the microwave signal (10--100\,GHz) to an observable level and allow one to
    scan over a broad axion mass range.
	A sensitivity to QCD axion models is conceivable
	with 80~disks of 1\,m$^2$ area contained in a $10$\,Tesla field.

\end{abstract}


\maketitle
%

\section{Introduction}
The nature of dark matter (DM) is one of the most enduring cosmological mysteries.
One prime candidate, the axion, arises from the Peccei--Quinn (PQ) solution to the strong CP problem, the absence of CP violation in quantum chromodynamics (QCD).
The CP violating QCD phase $\theta$ is effectively replaced by the axion field whose potential is minimal at $\theta=0$~\cite{Peccei:2006as,Kim:2008hd, Agashe:2014kda}. Thus $\theta$ dynamically relaxes towards zero regardless of its initial conditions, satisfying the
neutron electric dipole moment constraints $\theta\lesssim 10^{-11}$ \cite{Baker:2006ts}.

Tiny relic oscillations with a frequency given by the axion mass $m_a$ around $\theta=0$ persist, acting as cold DM \cite{Preskill:1982cy,Abbott:1982af,Dine:1982ah,Sikivie:2009fv,Kawasaki:2013ae}. If DM is purely axionic, its local
galactic density  $\rho_a=(f_a m_a)^2\theta_0^2/2\sim 300~{\rm MeV}/{\rm cm}^3$ implies
$\theta\sim \theta_0\cos(m_a t)$ at the Earth, with 
$\theta_0\sim 4\times 10^{-19}$. 
While these oscillations could be detected, the main challenge is to scan over a huge frequency range as $m_a$ is unknown.

However, cosmology can guide our search. Causality implies that at some early time $\theta$ is uncorrelated between patches of causal horizon size. We consider two cosmological scenarios depending on whether cosmic inflation happens after (A) or before (B) that time.

In Scenario~A, one patch is inflated to encompass our observable universe while smoothing $\theta$ to a
single initial value $\theta_{\rm I}$. The cosmic axion abundance depends on both $\theta_{\rm I}$ and $m_a$, so 
the DM density can be matched for any $m_a$ allowed by astrophysical bounds~\cite{Raffelt:2006cw}
for a suitable $\theta_{\rm I}$.

In Scenario~B, the axion abundance is given by the average over random initial conditions and the decay of accompanying cosmic strings and domain walls.
Freed from the uncertainty in the initial conditions, Scenario~B provides a concrete prediction $m_a\sim 100~\mu$eV~\cite{Hiramatsu:2012gg,Kawasaki:2014sqa}, although with some theoretical uncertainty~\cite{Fleury:2015aca}.
\begin{figure}[!]
\centering
\includegraphics[width=7cm]{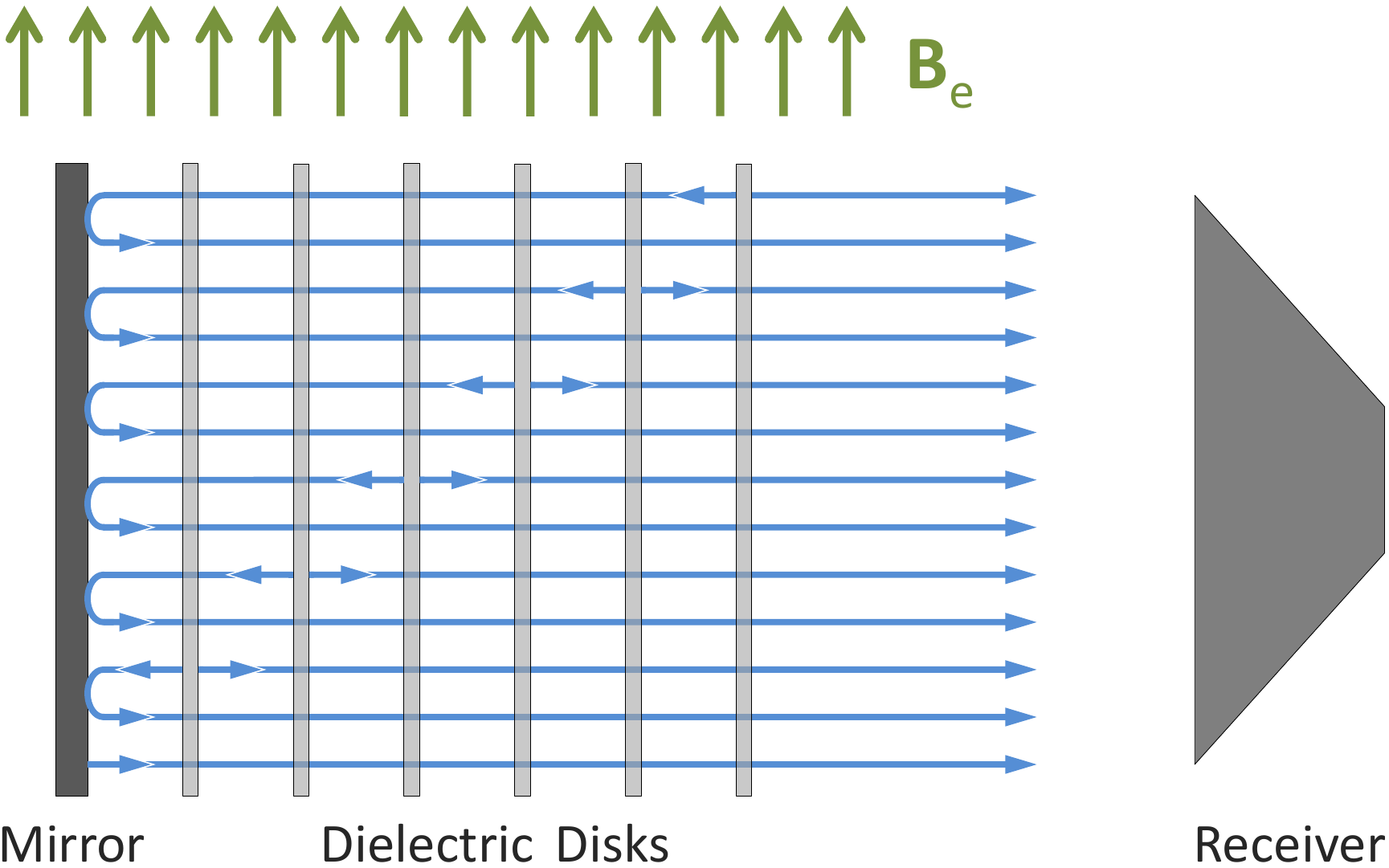}
\caption{A dielectric haloscope
consisting of a mirror and several dielectric disks placed in an external magnetic field ${\bf B}_{\rm e}$ 
and a receiver in the field-free region. A parabolic mirror (not shown) could be used to concentrate the emitted power into the receiver. Internal reflections are not shown.
}
\label{fig:LayeredDielectricHaloscope}
\end{figure}

Searches based on cavity resonators in strong magnetic fields (Sikivie's haloscopes~\cite{Sikivie:1983ip}) such as ADMX~\cite{Rybka:2014xca}, ADMX HF~\cite{Brubaker:2016ktl} or CULTASK~\cite{CULTASK} are optimal for $m_a\lesssim 10\,\mu\rm eV$. Much lower values of $m_a$ can be explored by nuclear magnetic resonance techniques like CASPER \cite{Budker:2013hfa} or with LC circuits~\cite{Sikivie:2013laa,Kahn:2016aff}.

The mass range favoured in Scenario~B is untouched by current experiments, and for cavity haloscopes will remain so for the foreseeable future. While fifth-force experiments~\cite{Arvanitaki:2014dfa} could search this region, they would not directly reveal the nature of DM. 
We present here a new concept to cover this important gap, 
capable of discovering \hbox{$\sim 100\,\mu$eV} mass axions. It consists of a series of parallel dielectric disks with a mirror on one side, all within a magnetic field parallel to the surfaces as shown in Fig.~\ref{fig:LayeredDielectricHaloscope}---a dielectric haloscope.

For large $m_a$ the greatest hindrance for conventional haloscopes is that the signal is proportional to the cavity volume $V$. With dimensions on the order of the axion Compton wavelength\footnote{We use natural units with $\hbar = c = 1$ and the Lorentz-Heaviside convention
$\alpha=e^2/4\pi$.} $\lambda_a= 2\pi/m_a$, $V\propto \lambda_a^3$ which decreases rapidly with $m_a$. While there are plans to couple multiple high-quality cavities, use open resonators, or compensate with extremely high magnetic fields and/or new detectors, these techniques may not prove practical for large $m_a$ \cite{CULTASK,Rybka:2014cya,Morris:1984nu,Lamoreaux:2013koa}.  

A radical approach to increase the volume is to use a dish antenna (i.e., a mirror) inside a $B$-field to convert axion DM into microwaves~\cite{Horns:2012jf}. 
The resonantly enhanced axion-photon conversion is then given up in favour of a large transverse area $A\gg \lambda_a^2$.

In our dielectric haloscope each disk
 emits electromagnetic (EM) waves similar to the mirror case.
 As the disks are semi-transparent, the waves emitted by each disk are reflected by and transmitted through the other disks before exiting. With suitable disk placement, these waves add coherently to the emitted power considerably with respect to a single mirror.
 While the use of $\lambda_a/2$-thick dielectric layers has already been proposed for cavity haloscopes~\cite{Morris:1984nu} and dish antennas~\cite{Jaeckel:2013eha},
 %
our disks do not have to be $\lambda_a/2$-thick because the coherence of the emitted waves can be controlled by the disk separations. 
Dielectric haloscopes can thus scan over a wide $m_a$ range
without needing to use disks with different thicknesses for each measurement.

\section{\boldmath Axion induced E-field} 

The axion-photon interaction is described by the Lagrangian density
\begin{equation}\label{lagrangian}
{\cal L_\text{int}} = 
-\frac{\alpha}{2\pi}C_{a\gamma}  {\bf E}\cdot {\bf B}\, \theta ,
\end{equation}
where ${\bf E}~\text{and}~{\bf B}$ are electric and magnetic fields, $\theta=a/f_a$
represents the axion field $a$,
 and $C_{a\gamma}={\cal E}/{\cal N}-1.92$ is an unknown parameter specific to each axion model, where ${\cal E}$ and ${\cal N}$ are the respective EM and colour anomalies of the PQ symmetry~\cite{Kim:2008hd}. 
We are interested in galactic DM axions, which are in a highly occupied state and very non-relativistic ($v_a\lesssim 10^{-3}$).
The associated de Broglie wavelength, $2\pi/m_a v_a\gtrsim 12.4\,{\rm m}\,(100\,\mu$eV$/m_a)$, is larger than the dimensions of our dielectric haloscope, a cubic meter. 
Thus we treat $\theta$ as a classical field and take the zero-velocity limit
which implies an (approximately)  homogeneous and monochromatic $\theta(t)\simeq \theta_0 \cos(m_a t)$.
The axion mass $m_a= 57.0\, {\rm \mu eV}\,\left(10^{11}{\rm\,GeV}/f_a\right)$ is related to the axion decay constant $f_a$~\cite{diCortona:2015ldu} which we treat as a free parameter.  

The interaction \eqref{lagrangian} enters as a current on the right-hand side (RHS) of the Amp\`ere-Maxwell equation,
\be
\nabla\times {\bf B}-\epsilon \dot{\bf E} = \frac{\alpha}{2\pi}C_{a\gamma} {\bf B}\, \dot \theta  ,
\ee
for a medium with permeability $\mu=1$ and dielectric constant $\epsilon$.
When a static and homogeneous external magnetic field ${\bf B}_{\rm e}$ is applied, the axion DM field sources a tiny electric field 
\begin{equation}
\label{eq:Ea}
{\bf E}_a(t)=- \frac{\alpha}{2\pi\epsilon}C_{a\gamma} {\bf B_{\rm e}}\,\theta(t)\, 
\end{equation}
which is discontinuous at the interface between media of different $\epsilon$. 
To satisfy the usual continuity requirements, ${\bf E}_{\parallel,1} ={\bf E}_{\parallel,2}$ and ${\bf B}_{\parallel,1} ={\bf B}_{\parallel,2}$ (recall $\mu=1$),
EM waves of frequency $\nu_a=m_a/2\pi$ must be present to compensate for the discontinuity.
In effect, breaking translational invariance couples the approximately non-propagating axion-induced $E$-field with propagating EM waves, which we aim to detect. The EM waves are emitted perpendicularly to the disk surfaces in our zero-velocity limit~\cite{Horns:2012jf}. We require flat disks of large area to avoid diffraction, $A\gg \lambda_a^2$, so we can work in a one-dimensional (1D) framework. For maximum effect, ${\bf E}_a$---and thus ${\bf B}_{\rm e}$---must be parallel to the disk surfaces~\cite{Horns:2012jf}.

The gain from a dielectric haloscope with respect to a mirror alone
can be quantified in terms of a boost factor $\beta$.
Comparing the $E$-field inside the mirror ($E\simeq 0$) 
with the axion-induced one in vacuum 
$E_0\equiv \alpha/(2\pi)\, | C_{a\gamma} {\bf B}_{\rm e}\,\theta_0|$, one finds that the continuity at the surface requires the emission of an EM wave of amplitude $E_0$ with a power per unit area $P_{0}/A=E_0^2/2$~\cite{Horns:2012jf}.
With additional dielectric disks as shown in Fig.~\ref{fig:LayeredDielectricHaloscope},
the axion field induces plane EM waves for every change of media.
By tuning the disk separations, constructive interference can enhance the amplitude of the emitted EM wave to $E_{\rm out}$ at the receiver.
The boost factor is thus defined as
\be
\beta(\nu_a) \equiv |E_{\rm out}(\nu_a)/E_0| .
\label{eq:boost}
\ee
With this definition, and $E_0$ as introduced above, the resulting output power of the dielectric haloscope 
per unit area is 
\begin{eqnarray}
\frac{P}{A} = \beta^2\, \frac{P_{0}}{A}
= 2.2\times 10^{-27}\, \frac{\rm W}{\rm m^2}\,\beta^2
\(\frac{B_{\rm e}}{10~{\rm T}}\)^2 C_{a\gamma}^2 \, .
\label{eq:PoverAdh}
\end{eqnarray}
The boost factor is
calculated by matching 
${\bf E}_a$ in each region (dielectric disk or vacuum) with left and right-moving EM waves as imposed by the continuity of the total ${\bf E}_{||}$ and ${\bf B}_{||}$ fields at the interfaces. 
A transfer matrix formalism has been developed for this task~\cite{Millar:2016cjp}.

The desired enhancement, $\beta\gg 1$, comes from two effects, which generally act together but can be differentiated in limiting cases. 
These effects depend on the optical thickness of one disk $\delta=2\pi \nu d \sqrt{\epsilon} $, with $d$ the physical thickness and $\nu$ the frequency, which sets the transmission coefficient of a single disk, found to be \hbox{${\cal T}=i 2 \sqrt{\epsilon}/[i2\sqrt{\epsilon}\cos\delta+(\epsilon+1)\sin\delta]$}. 
When $\delta=\pi,3\pi,5\pi,...$, the disk is transparent (${\cal T}=0$) and the emission from different disks can be added constructively by placing them at the right distance. 
When ${\cal T}\neq 0$, the spacings can be adjusted to form a series of leaky resonant cavities where $E$-fields are boosted by reflections between the disks. In general both the simple sum of emitted waves and resonant enhancements are important.

\section{ Comparison with cavity haloscopes} 
More insight is gained by contrasting dielectric haloscopes with resonant cavities. The signal power extracted from a cavity haloscope resonantly excited by axion DM is usually expressed as~\cite{Sikivie:1985yu}
\be
P_{\rm cav} = \kappa {\cal G} V \frac{Q}{m_a} \rho_{a} g_{a\gamma}^2B_{\rm e}^2, 
\label{eq:soverlap}
\ee
where $V$ is the cavity volume, $g_{a\gamma}=-\alpha/(2\pi f_a)C_{a\gamma}$, 
$Q$  
the loaded quality factor, and $\kappa$ the ratio of signal power to total power loss. The geometry factor 
\hbox{${\cal G}= \(\int dV {\bf E}_{\rm cav}\cdot {\bf B}_{\rm e}\)^2/(V B_{\rm{e}}^2\int dV {\bf E}_{\rm cav}^2)$}, 
is the normalised overlap integral of $B_{\rm e}$ with the $E$-field of the resonant mode, ${\bf E}_{\rm cav}({\bf x})$, which is calculated by imposing closed boundary conditions. 

In contrast, dielectric haloscopes are open systems, which are in general non-resonant.  
Nevertheless, after lengthy algebraic manipulations one can express the boost factor~\eqref{eq:boost}
as an overlap integral~\cite{Millar:2016cjp}, 
\be
\beta \simeq \frac{m_a}{2 B_{\rm e} E_0}\left |\int dx E_{\rm dh} B_{\rm e}\right |.
\ee
With this expression, Eq.~\eqref{eq:PoverAdh} becomes
highly reminiscent of Eq.~\eqref{eq:soverlap}.  
This can actually be turned into a dictionary: the power output of a dielectric haloscope can be calculated with Eq.~\eqref{eq:soverlap} 
by defining
\begin{eqnarray}
Q_{\rm dh}\equiv\frac{1}{4}\frac{\int dx |E_{\rm dh}|^2}{E_0^2/m_a}, \quad 
{\cal G}_{\rm dh}&\equiv&\frac{\left|\int dx E_{{\rm dh}} B_{{\rm e}}\right |^2}{L B^2_{{\rm e}} \int dx |E_{\rm dh}|^2}, 
\label{colindro}
\end{eqnarray}
where $L$ is the length of the haloscope. Note that $\kappa$ is implicitly contained in ${\cal G}_{\rm dh}~\text{and}~ Q_{\rm dh}$. 
Unlike for a cavity, $E_{{\rm dh}}(x)$ is defined as the $E$-field inside the dielectric haloscope
obtained by shining in an EM 
wave of amplitude $E_0$ from the RHS in Fig.~\ref{fig:LayeredDielectricHaloscope}. 
This is a non-trivial conceptual change as the physical axion-induced $E$-field
(obtained with the transfer matrix formalism) is in general different to $E_{{\rm dh}}(x)$~\cite{Millar:2016cjp}. However, when we consider highly resonant dielectric haloscopes (for instance by choosing a very reflective rightmost disk in Fig.~\ref{fig:LayeredDielectricHaloscope})
the definitions of $E_{\rm cav}$ and $E_{\rm dh}$ coincide as expected.  

Equation~\eqref{colindro} suggests the interpretation of $\beta^2$ as $m_a L {\cal G}_{\rm dh} Q_{\rm dh}$. Our proposal is to enhance $P$ by increasing
$V_{\rm dh}\equiv L A$, ${\cal G}_{\rm dh}$ and $ Q_{\rm dh}$ with appropriately placed disks. This strategy sounds similar to enhancing $V$ and ${\cal G}$ of a strictly resonant cavity by either modifying the magnetic field~\cite{Rybka:2014cya} or using dielectrics~\cite{Morris:1984nu,Rybka2016Jeju}.
However, our dielectric haloscope would be a poor resonator, compensating a relatively low $Q_{\rm dh}$ with a huge
$V_{\rm dh}$. 
This has completely different engineering challenges than an intermediate volume, high-$Q$ cavity. 

Large-$V$ resonators are typically more complicated mechanically, which implies longer tuning times. 
For high-$Q$ cavities, as $V$ increases modes tend to clutter~\cite{Baker:2011na} and it becomes increasingly difficult to identify and tune to them. Mode crossings become more frequent, with concomitant forbidden frequencies. 
Furthermore, for a \hbox{large-$V$} resonant cavity the coupling of the output port needs to increase locally to compensate for a longer time-of-flight of photons in the cavity, leading to stronger mode distortions. 

In our case, we have the flexibility to compensate for longer tuning times by using broadband and non-resonant configurations not possible with cavity haloscopes. Further, we avoid mode crossings by not having resonant modes, rather quasi-modes with very broad widths.
Lastly, by making the system as 1D as possible with a large $A$, and extracting the signal homogeneously across the last dielectric disk 
before the receiver, we minimise any distortion caused by a local coupling to the detector. 
This allows us to avoid some of the issues of large-$V$ resonators.

\section{Properties of the boost factor} 
After choosing $\epsilon$ and the thickness of the disks $d$, the distances between disks remain as the only free parameters of our dielectric haloscope,  still leaving considerable control over the frequency response. We will actually be interested in two types of configurations: one with a flat broadband response to measure a large frequency range in one go, and another with a larger $\beta$ over a narrow band to discard statistical fluctuations and do precision axion physics in case of a discovery. We can numerically generate such configurations.

One can predict the general behaviour of $\beta$ by using what we call the Area Law: $\int \beta^2 d\nu_a$
is proportional to the sum over interfaces, which holds exactly when integrating over \hbox{$0\leq\nu_a \leq \infty$}, and is a good approximation for frequency ranges containing the main peak~\cite{Millar:2016cjp}. 
According to the Area Law, an increase in the number of disks gives a linear increase in $\int \beta^2 d\nu_a$. For a single set of disks $\int \beta^2 d\nu_a$ is constant; one can trade width for power and vice versa, but cannot gain in both simultaneously.

\begin{figure}[t]
\begin{center}
\includegraphics[width=8.5cm]{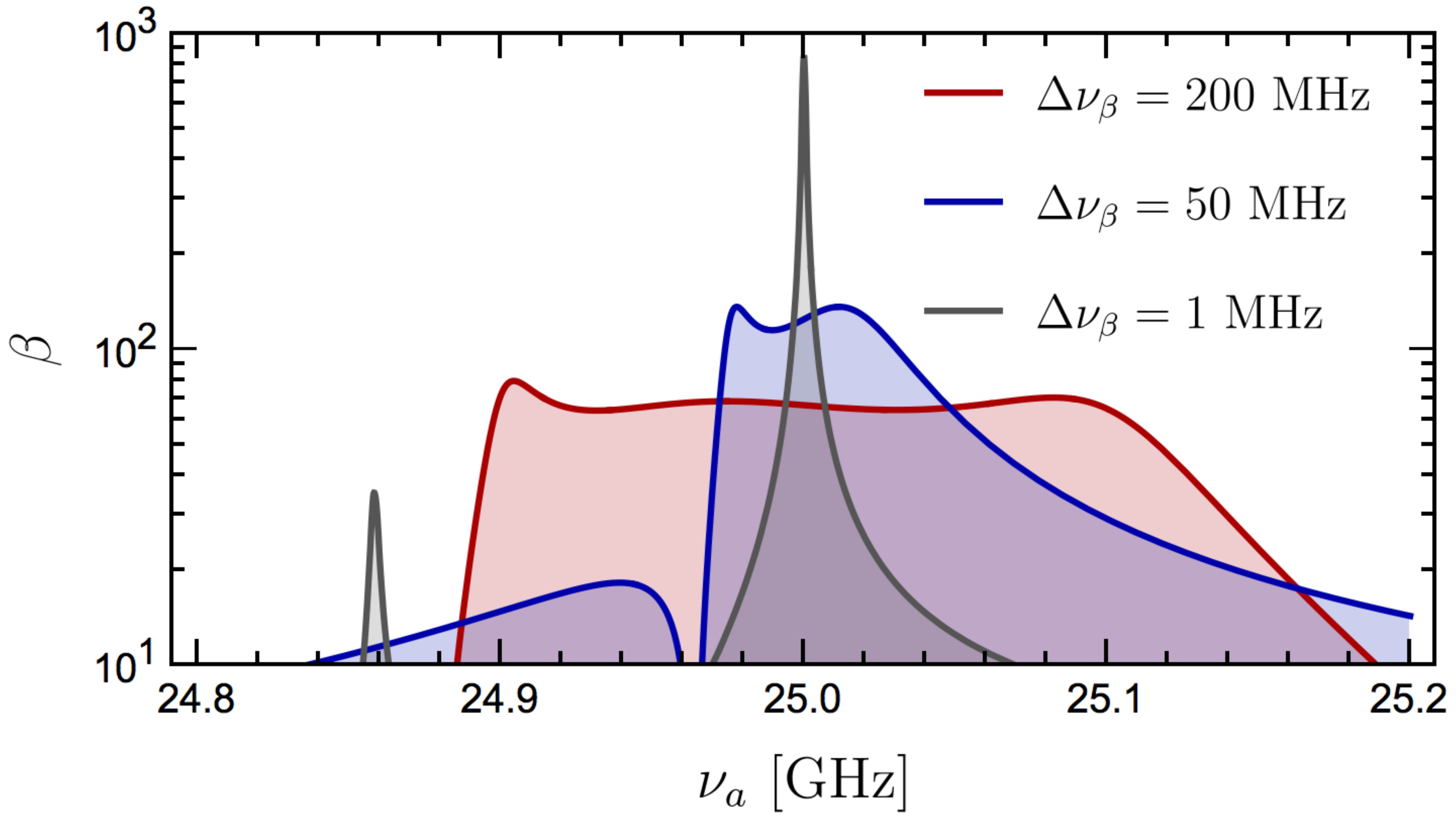}
\caption{Boost factor $\beta(\nu_a)$ for configurations optimised for $\Delta\nu_\beta=200,\, 50\,\,{\rm and}\,\,1~\rm{MHz}$ (red, blue and grey) centred on $25\,{\rm GHz}$ using a mirror and 20~dielectric disks ($d=1\,{\rm mm}$, $\epsilon=25$). }
\label{fig:bandwidths}
\end{center}
\end{figure}

Figure~\ref{fig:bandwidths} depicts $\beta(\nu_a)$ for a dielectric haloscope consisting of a mirror and 20 disks (\hbox{$d=1~{\rm mm}$}, \hbox{$\epsilon=25$}). Spacings have been selected to maximise the minimum boost factor $\beta_{\rm min}$ within three bandwidths \hbox{$\Delta\nu_\beta=1,\,50\,\,\text{and}\,\,200~{\rm MHz}$} centred on $25~{\rm GHz}$ (our benchmark frequency corresponding to $m_a= 103.1~\mu$eV). 
These curves illustrate the Area Law: as we vary the bandwidth, the power changes by roughly the same factor. 
While the area is conserved, it appears to be more effectively used in broadband configurations where we were able to obtain flatter $\beta(\nu_a)$ curves.  
Using the Area Law we extrapolate to the 80~disk setup described below, reaching $\beta_{\rm min}\sim275$ across 50~MHz. The achievable $\beta_{\rm min}$ changes with the optical thickness of a disk---in the example shown $\delta\sim 0.8\pi$, neither transparent nor fully reflective. Note that for a given $d$, there are frequency bands around $\nu_a=1,3,5,...\times\pi/(\sqrt{\epsilon}d)$ for which the disks are transparent and $\beta$ is limited to the sum of the EM waves $\beta_{\rm min}\leq 2N+1$, where $N$ is the number of disks. Further, at $\nu_a=2,4,6,...\times\pi/\sqrt{\epsilon}d$ the disks do not emit any radiation. Thus when large $\beta$ or flexibility is required one must avoid these frequencies with a different set of disks.

\section{Setup}
Dielectrics with large $\epsilon$ enhance emission and resonance effects, leading to a higher $\beta$. A good candidate could be LaAlO$_3$ ($\epsilon\sim 25,\, \mu\sim 1$), which has a very small loss tangent $\sim 10^{-5}$ at temperatures below $\sim 80$~K. The transverse area is limited by the feasibility and cost of an intense magnet: apertures larger than one square meter are extremely challenging.
Dielectric disks of that area would be made by tiling smaller pieces and could be repositioned using precision motors. The precision required can be estimated by studying the analytically tractable cases---the single cavity and the $\lambda_a/2$ symmetric case~\cite{Jaeckel:2013eha}. 
We find that the positioning errors have to be kept below $\sim 200~\mu {\rm m} \sqrt{10^2/\beta}\({100~\mu\rm eV}/{m_a}\)$. 
For example, for $\beta\sim 10^3$,
$\sigma\ll 60~\mu$m would be needed for $m_a=100~\mu$eV, though the exact sensitivity depends on the configuration. Thus it will probably be more practical to use a broadband search technique, scanning a larger mass range in each longer measurement. We are investigating this requirement with a prototype setup using 20~cm diameter Al$_2$O$_3$ disks. Comparisons of the simulated and measured transmissivity and reflectivity (which are correlated with $\beta$) indicate that few $\mu{\rm m}$ precision could be achieved. Diffraction appears to be negligible for the setup, but full numerical studies are required.

Thermal emission of the disks and mirror contributes to the noise of the experiment, but is suppressed when compared to a black body by both small dielectric losses and a good reflectivity, respectively. However, the haloscope will reflect thermal emission from the detector into itself. This effect can be reduced by cooling the detector to cryogenic temperatures. As detector, we would use a broadband corrugated horn coupled to a linear amplifier like state-of-the-art
high-electron-mobility transistor (HEMT)
for its broadband capabilities (operable up to 40 GHz) and/or quantum limited amplifiers. Using a HEMT detector at room temperature we measured a $10^{-21}$ W signal at 17 GHz at $ 6\sigma$ in a one week run---we expect 100 times better performance at cryogenic temperatures. With $\beta\sim 400$ we would be sensitive to QCD axions.

\section{Discovery Potential}
For a practical experiment one must scan across $m_a$. The procedure consists of arranging the disks to achieve a roughly constant $\beta$ in a region $\Delta\nu_\beta$, measuring for some time $\Delta t$, and readjusting the distances to measure an adjacent frequency range. 
The required $\beta$ for a reasonably short $\Delta t$ is given by Dicke's radiometer equation for the desired signal to noise ratio,
$S/N= (P/T_{\rm sys})\sqrt{\Delta t/\Delta \nu_a}$, \exclude{\label{eq:dicke}}where the system noise temperature is $T_{\rm sys}$ and the axion line width $\Delta \nu_a\sim 10^{-6}\nu_a$. 
Collecting the signal with efficiency $\eta$, we get 
\begin{eqnarray}
\frac{\Delta t}{1.3\,\rm days}&\sim&
\(\frac{{\rm S}/{\rm N}}{5}\)^2\(\frac{400}{\beta}\)^4
\(\frac{1\,{\rm m}^2}{A}\)^2\(\frac{m_a}{100~\mu{\rm eV}}\)\nonumber\\
&&\times\(\frac{T_{\rm sys}}{8\,\rm K}\)^2
\(\frac{10\,\rm T}{B_{\rm e}}\)^4\(\frac{0.8}{\eta}\)^2C_{a\gamma}^{-4} \,. \label{eq:scan2}
\end{eqnarray}
In a single measurement we simultaneously search $\Delta\nu_\beta/\Delta\nu_a$ possible axion ``channels". Thus the time to scan a given frequency range scales inversely to $\beta^4\Delta\nu_\beta$. As the Area Law implies that $\beta^2\Delta\nu_\beta$ is approximately conserved, it appears that narrow resonant peaks are optimal, as with conventional cavity haloscopes. 

However, both the required placement precision and the time $t_{\rm R}$ needed to reposition the disks limit $\beta$. With 80 disks to be adjusted, $t_{\rm R}$ will be non-negligible---for an optimal scanning rate the measurement and readjustment times must be similar. 
The value of $\beta$ required to reach $|C_{a\gamma}|\sim 1$ increases with $m_a$, so the dielectric haloscope could be adjusted to scan a wider mass range simultaneously at low masses and increase $\beta$ decreasing the bandwidth as the scan proceeds to higher $m_a$. If a potential signal is found, the dielectric haloscope could be reconfigured to enhance $\beta$ at that frequency, quickly confirming or rejecting it at high significance. 

\begin{figure}[bt]
\centering
\includegraphics[width=8.3cm]{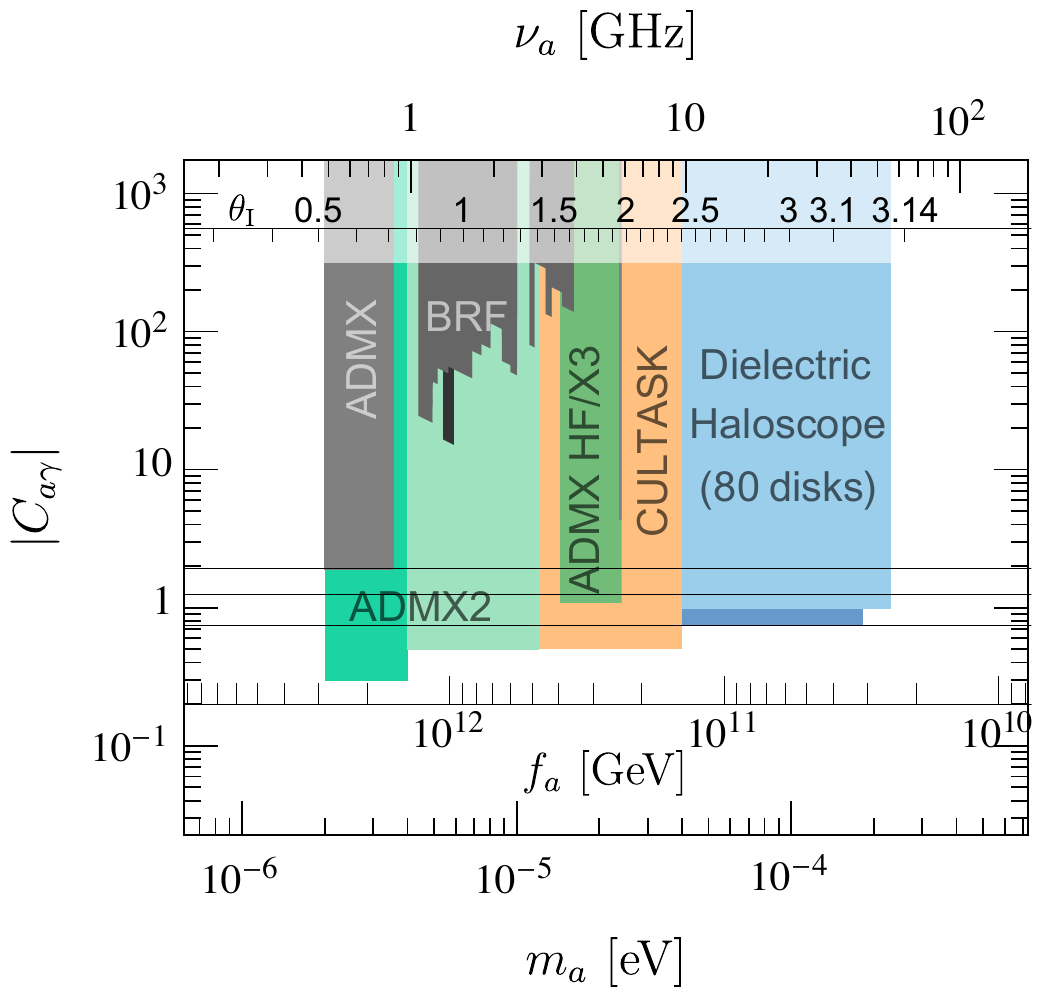}
\caption{Two examples of the discovery potential (light and dark blue) of our dielectric haloscope using 80 disks ($\epsilon=25$, $A=1\,{\rm m}^2$, $B_{\rm e}=10\,{\rm T}$, $\eta=0.8$, $t_{\rm R}=1\,{\rm day}$) with quantum limited detection in a 3-year campaign. We also show exclusion limits (gray) and sensitivities (coloured) of current and planned cavity haloscopes~\cite{Brubaker:2016ktl,Asztalos:2009yp,Hagmann:1990tj,Wuensch:1989sa,Carosi2016Jeju,vanBibber2015Zaragoza,CULTASK}. 
The upper inset shows the initial angle $\theta_{\rm I}$ required in Scenario A~\cite{Borsanyi:2016ksw}. The lower inset depicts the $f_a$ value corresponding to a given $m_a$, and the three black lines denote $|C_{a\gamma}|=1.92,1.25,0.746$.
Note that Scenario B predicts $ 50~\mu{\rm eV}\lesssim m_a\lesssim 200~\mu{\rm eV}$~\cite{Kawasaki:2014sqa,Ballesteros:2016euj}.} 
\label{fig:reach}
\end{figure}

In Fig.~\ref{fig:reach} we show the discovery potential of an 80 disk experiment with a run time of three years, by extrapolating from our $25~{\rm GHz}$ solutions using the Area Law. We have used $A=1\,{\rm m}^2$, $\epsilon= 25$, and  $B_{\rm e}=10\,{\rm T}$. We assume 80\% detection efficiency, quantum limited detection ($T_{\rm sys}\sim m_a$) and a conservative $t_{\rm R}=1$\,day. We give two examples, reaching $|C_{a\gamma}|=1$ and $0.75$ in light blue and dark blue, respectively. Similar results can be achieved by using a two stage process. A five year run with commercially available HEMT amplifiers with $T_{\rm sys}=8\,{\rm K}$ would cover the low-mass range, for example $m_a\lesssim 120 \,\mu{\rm eV}$ with $|C_{a\gamma}|=1$. The high-mass range $m_a\lesssim 230\, \mu{\rm eV} $ shown in Fig.~\ref{fig:reach} would require another two years with a quantum limited detector. Adding disks and extending the run time could expand the search to $m_a\lesssim 400\,\mu$eV.

Our haloscope would cover a large fraction of the high-mass QCD axion parameter space with sensitivity to $|C_{a\gamma}|\sim 1$. In Scenario A, these masses correspond to large, but still natural, initial angles $2.4\lesssim \theta_{\rm I}\lesssim 3.12$~\cite{Borsanyi:2016ksw}. Scenario B, our main goal, would be covered including the theoretical uncertainty in $m_a$ ($50\text{--}200~\mu$eV~\cite{Kawasaki:2014sqa,Ballesteros:2016euj}) for KSVZ-type models with short-lived domain walls (${\cal N}=1$). Prime examples include the recent SMASH$_{d,u}$ models ($ {\cal E}=2/3,8/3$)~\cite{Ballesteros:2016euj}. Models with ${\cal N}>1$ require $m_a\gtrsim 1~{\rm meV}$~\cite{Kawasaki:2014sqa,Ringwald:2015dsf}, beyond our search range. However, with some exceptions~\cite{Lazarides:1982tw} these models generally require a tuned explicit breaking of the PQ symmetry to avoid a domain-wall dominated universe~\cite{Sikivie:1982qv}.

\section{Conclusion}

In this Letter we have proposed a new method to search for high-mass ($40\text{--}400~\mu$eV) axions by using a mirror and multiple dielectric disks contained in a magnetic field---a dielectric haloscope. The key features are a large transverse area and the flexibility to use both broadband and narrow-band search strategies. With 80 disks one could search a large fraction of this high-mass range with sensitivity to the QCD axion. 

\begin{acknowledgements}
\section*{Acknowledgments}
We acknowledge partial support by the Deutsche Forschungsgemeinschaft
through Grant No.\
EXC 153 (Excellence Cluster ``Universe''), the
Alexander von Humboldt Foundation,
as well as the European Commission under ERC Advanced Grant 339169 and
under the Innovative
Training Network ``Elusives'' Grant No.\ H2020-MSCA-ITN-2015/674896. J.R.\ is supported by the Ramon y Cajal Fellowship 2012-10597 and FPA2015-65745-P (MINECO/FEDER). Part
of this work
was performed at the Bethe Forum ``Axions'' (7--18 March 2016),
Bethe Center for Theoretical Physics,
University of Bonn, Germany.
\end{acknowledgements}

\end{document}